\documentclass{PoS}
\usepackage{placeins}

\title{Towards Lefschetz thimbles regularization of heavy-dense QCD}

\ShortTitle{Towards Lefschetz thimbles regularization of heavy-dense QCD}

\author{\speaker{Kevin Zambello} and Francesco Di Renzo\\
        Dipartimento di Scienze Matematiche, Fisiche e Informatiche, Universit\`a di Parma and INFN, Gruppo Collegato di Parma, I-43124 Parma, Italy\\
        E-mail: \email{kevin.zambello@pr.infn.it}, \email{francesco.direnzo@pr.infn.it}}

\abstract{
At finite density, lattice simulations are hindered by the well-known sign problem: for finite chemical potentials, the QCD action becomes complex and the Boltzmann weight $e^{-S}$
cannot be interpreted as a probability distribution to determine expectation values by Monte Carlo techniques. Different workarounds have been devised to study the QCD phase diagram,
but their application is mostly limited to the region of small chemical potentials. The Lefschetz thimbles method takes a new approach in which one complexifies the theory and
deforms the integration paths. By integrating over Lefschetz thimbles, the imaginary part of the action is kept constant and can be factored out, while $e^{-Re(S)}$ can be interpreted
as a probability measure. The method has been applied in recent years to more or less difficult problems. Here we report preliminary results on Lefschetz thimbles regularization of
heavy-dense QCD. While still simple, this is a very interesting problem. It is a first look at thimbles for QCD, although in a simplified, effective version. From an algorithmic point
of view, it is a nice ground to test effectiveness of techniques we developed for multi thimbles simulations.
}

\FullConference{The 36th Annual International Symposium on Lattice Field Theory - LATTICE2018\\
		22-28 July, 2018\\
		Michigan State University, East Lansing, Michigan, USA.}

\begin{document}

\section{Introduction}
The $T-\mu_B$ phase diagram of QCD is a subject of utmost importance, due to its relevance to many physical systems, but its exploration by lattice calculations
is a difficult task at finite baryon chemical potential, where the fermionic term of the action becomes complex, hindering Monte Carlo simulations.
This issue is known as the sign problem. Lefschetz thimbles regularization \cite{cristoforetti,kikukawa} is one of the many approaches to attack the sign problem.
This relatively new technique has been applied to various toy models and, recently, to one-dimensional QCD \cite{eruzzi2,felix}.

The main idea of thimble regularization is to complexify the degrees of freedom of the theory and to compute expectation values as sums of integrals over manifolds
attached to the critical points of the complexified theory,
\begin{equation}
\langle O \rangle = \frac{1}{Z} \int d^n x \thinspace O(x) \thinspace e^{-S(x)} = \frac{1}{Z} \sum_{\sigma} n_\sigma e^{-iS^I_\sigma} \int_{\mathcal{J}_\sigma} d^n z 
\thinspace O(z) \thinspace e^{-S^R(z)} \mbox{ .}
\label{eq:O}
\end{equation}
In eq. \ref{eq:O} the index $\sigma$ labels the critical points $z_\sigma$ ($z = x + i \thinspace y$ is a shorthand notation for the complexified fields), $n_\sigma$
are integer coefficients and the manifolds $\mathcal{J}_\sigma$, called thimbles, are defined as the union of all the steepest ascent paths originating from the critical
point $z_\sigma$, i.e. the solutions of the steepest ascent (SA) equations
$$\frac{dz_i}{dt} = \frac{\partial \overline{S(z)}}{\partial \overline{z_i}}\mbox{ ,\hspace{2cm}}z_i(-\infty) = z_{\sigma,i} \mbox{ .}$$
Along the flow, the imaginary part of the action stays constant, which ensures that the terms $e^{-iS^I_\sigma}$ can be factored out.
The real part of the action is always increasing, which ensures the convergence of the integrals over the thimbles. It should be noted that
in general the integration measure $dz^n$ is not parallel to the tangent space $T_z \mathcal{J}_\sigma$ at the point $z$; after a change of
coordinates from the canonical basis to a basis of the tangent space, the integration measure becomes $d^n z = detV_\sigma \thinspace d^n y = |det V_\sigma| 
\thinspace e^{i \omega_\sigma} d^n y$, where $V_\sigma$ is the matrix having as columns the basis vectors $\{ V_\sigma^{(i)} \}$ of the tangent space at the point $z$.
The residual phase $e^{i \omega_\sigma}$ could in principle give rise to a residual sign problem, though so far this has been observed to be much milder
than the original sign problem.

The tangent space at a critical point $z_\sigma$ is spanned by the Takagi vectors, which can be obtained by solving the Takagi problem
$$H(z_\sigma) v^{(i)} = \lambda^{(\sigma)}_i \overline{v^{(i)}}\mbox{, }$$
where $H_{i,j}(z_\sigma) = \frac{\partial^2 S}{\partial z_i \partial z_j} (z_\sigma)$ is the Hessian of the complexified action, $\{v^{(i)}\}$ are the Takagi
vectors and $\{\lambda^{(\sigma)}_i\}$ their associated Takagi values.
A point $z$ on the thimble can be pinpointed by the direction $\hat{n}$ taken by the flow at the critical point ($\hat{n}$ belongs to the tangent space)
and the integration time $t$.
By using this parametrization and introducing a partial partition function $Z_{\hat{n}} = ( 2 \sum_i \lambda_i^{(\sigma)} n_i^2 ) \int dt \thinspace 
e^{-S_{eff}(\hat{n}, t)}$ (the effective action is defined as $S_{eff}(\hat{n}, t) \equiv S_R(\hat{n}, t) - ln(|det \thinspace V_\sigma(\hat{n}, t)|)$),
expectation values can be rewritten using the multi thimbles decomposition \cite{eruzzi2}
\begin{equation}
\langle O \rangle = \frac{\sum_\sigma n_\sigma Z_\sigma e^{-iS^I_\sigma} \langle O e^{i\omega}\rangle_\sigma}{\sum_\sigma n_\sigma Z_\sigma e^{-iS^I_\sigma} \langle e^{i\omega}\rangle_\sigma}\mbox{ .}
\label{eq:thimble_decomposition}
\end{equation}
Here:
\begin{itemize}
\item $f_{\hat{n}} = \frac{1}{Z_{\hat{n}}} ( 2 \sum_i \lambda_i^{(\sigma)} n_i^2 ) \int dt \thinspace f (\hat{n}, t) e^{-S_{eff}(\hat{n}, t)}$
      is a term which looks like an average of the observable $f (\hat{n}, t)$ over an entire SA path.
\item $\langle f \rangle_\sigma = \int D\hat{n} \thinspace \frac{Z_{\hat{n}}}{Z_\sigma} \thinspace f_{\hat{n}}$
      is a term which looks like a functional integral over all SA paths. This allows to introduce an importance sampling over thimbles
      by extracting directions $\hat{n} \propto \frac{Z_{\hat{n}}}{Z_\sigma}$.
\item $Z_\sigma = \int D\hat{n} \thinspace Z_{\hat{n}}$ is a term which weights the contribution of different thimbles.
\end{itemize}
Since the tangent space at a generic point $z$ is in general unknown, the evaluation of $det \thinspace V_\sigma(\hat{n}, t)$ requires to transport the basis of the tangent space 
all the way along the flow. This asks for solving the parallel transport (PT) equations, which must be integrated numerically alongside the SA equations.

\section{Towards regularization of heavy-dense QCD}

\subsection{The theory}
In this work we approach heavy-dense QCD by the Lefschetz thimbles method. This theory has been investigated in the past by the Complex Langevin method
both using the full gauge action and a hopping parameter expansion \cite{aarts} and by the means of an effective formulation that can be obtained from QCD by a
combined strong-coupling and hopping parameter expansion. The latter is a 3d effective theory, whose only degrees of freedom are
the Polyakov loops, governed by the action \cite{philipsen2,philipsen1,philipsen3}
$$S = S_G + S_F = -\lambda \sum_{<x,y>} \left(TrW_x TrW_y^\dagger + TrW_x^\dagger TrW_y\right)$$
      $$- 2 \sum_x ln \left( 1 + h_1 TrW_x + h_1^2 TrW_x^\dagger + h_1^3 \right)$$
      $$+ 2 h_2 \sum_{<x,y>}  \left( \frac{h_1 TrW_x + 2 h_1^2 TrW_x^\dagger + 3 h_1^3}{1 + h_1 TrW_x + h_1^2 TrW_x^\dagger + h_1^3} \right)
                                     \left( \frac{h_1 TrW_y + 2 h_1^2 TrW_y^\dagger + 3 h_1^3}{1 + h_1 TrW_y + h_1^2 TrW_y^\dagger + h_1^3} \right) + O(k^4) \mbox{ ,}$$
with
      $$\lambda = u^{N_t} e^{N_t \cdot (4u^4+12u^5+\ldots)} \mbox{ , \hspace{0.5cm}}
      h_1 = (2 k e^{\mu})^{N_t}[1 + 6k^2 N_t \frac{u-u^{N_t}}{1-u} + \ldots] \mbox{ , \hspace{0.5cm}} h_2 = k^2 \frac{N_t}{3}[1 + 2 \frac{u-u^{N_t}}{1-u} + \ldots]$$
where $u \approx \frac{\beta}{18}$, $k$ is the hopping parameter and $W_x = \prod_{t=1}^{N_t} U_0(x, t)$ is the Polyakov loop.
The first sum extends over the $N_{sites}$ spatial sites of the lattice, while the second sum extends over nearest neighbors.

This theory features a mild sign problem and it has been used to investigate heavy-dense QCD in the cold regime \cite{philipsen1},
where $N_t \gg 1$, $\lambda \approx 0$ and the gauge term of the action becomes numerically negligible.
This is also an interesting study to carry out by thimble regularization for two reasons. On one hand, 
it allows to investigate the interplay between the spatial extension of the lattice and the number of relevant critical points.
On the other hand, under the approximation of heavy quarks, it allows to explore a corner of the QCD phase diagram, the one in which $T \approx 0$ $MeV$ and 
$\mu \approx \mu_c \equiv m = - ln (2k)$, where the onset of cold nuclear matter takes place.
Here we make the very first step, in which one neglects the $O(k^2)$ term.
This simplification amounts to taking the limit of very heavy quarks and, though observables in this simplified theory are independent from the volume,
their value is the result of a volume dependent recombination of contributions from different thimbles emerging from the thimble decomposition in eq. \ref{eq:thimble_decomposition}.
Therefore the theory remains a nice playground to test the effectiveness of the techniques we developed for multi thimbles simulations.

\subsection{Thimble regularization}
We choose to work in a convenient gauge, where $W_x = U_x \equiv U_0(x,t=0)$. Hence the complexified action becomes

$$S = - 2 \sum_x ln \left( 1 + h_1 TrU_x + h_1^2 TrU_x^{-1} + h_1^3 \right) \mbox{ .}$$

\noindent The gradient of the action determines the SA equations for the fields:
\begin{eqnarray}
\nabla^a_x S [U] & = & -2i \frac{h_1 Tr[T^a U_x] - h_1^2 Tr[U_x^{-1} T^a]}{1 + h_1 Tr U_x + h_1^2 Tr U_x^{-1} + h_1^3} \nonumber \\
\frac{dU_x}{dt} & = & i (\sum_a T^a \overline{\nabla^a_x S [U]}) U_x \mbox{ .}
\label{eq:SA}
\end{eqnarray}

\noindent After introducing the notation $(M)_{Tr}$ for the traceless part of the matrix $M$ and after defining $\alpha_x \equiv (1 + h_1 Tr U_x + h_1^2 Tr U_x^{-1} + h_1^3)^{-1}$,
the drift can be explicitly written as
$$\sum_a T^a \overline{\nabla^a_x S[U]} = \left(-i \alpha_x (h_1 U_x - h_1^2 U_x^{-1})_{Tr} \right)^\dagger \mbox{.}$$

\noindent By requiring a vanishing gradient, one finds that the critical points are mixtures of center elements of $SU(3)$,
$$\{ U_x = I_{3 \times 3} \thinspace e^{i \omega_x} \thinspace | \thinspace \omega_x \in \{ 0, \pm \frac{2 \pi}{3} \} \} \mbox{.}$$

\noindent The PT equations for the tangent space basis vectors read
\begin{equation}
\frac{dV_{zc}}{dt} = \sum_{w,a} \overline{V_{wa} \nabla^a_w \nabla^c_z S [U]} + \sum_{a,b} f^{abc} V_{za} \overline{\nabla^b_z S [ U ]} \mbox{ .}
\label{eq:PT}
\end{equation}
These are given in terms of the Hessian
$$\nabla^b_w \nabla^a_z S [U] = 2 \delta_{wz} \left[ \frac{(h_1 Tr [T^a T^b U_z] + h_1^2 Tr [U_z^{-1} T^b T^a])}{(1 + h_1 Tr U_z + h_1^2 Tr U_z^{-1} + h_1^3)} - \right.$$
$$\left. \frac{(h_1 Tr [T^a U_z] - h_1^2 Tr [U_z^{-1} T^a]) (h_1 Tr [T^b U_z] - h_1^2 Tr [U_z^{-1} T^b])}{(1 + h_1 Tr U_z + h_1^2 Tr U_z^{-1} + h_1^3)^2} \right] \mbox{.}$$

\noindent By "diagonalizing" the Hessian, one finds $8 \cdot N_{sites}$ Takagi vectors and values
$$v_{az}^{(j_1 j_2)} = \delta_{a j_1} \delta_{z j_2} e^{-i \frac{\phi(z)}{2}}$$
$$\lambda^{(j_1 j_2)} = | \gamma(j_2) |$$
with $$\gamma(z) = \frac{h_1 e^{i \omega_z} + h_1^2 e^{-i \omega_z}}{1 + 3 h_1  e^{i \omega_z} + 3 h_1^2  e^{-i \omega_z} + h_1^3} \equiv |\gamma(z)| \thinspace e^{i \phi(z)}$$
where the indexes $j_1$, $j_2$ used to label the vectors run respectively from $1 \ldots 8$ and $1 \ldots N_{sites}$.


\subsection{Critical points and semiclassical approximation}
The critical points of the theory are mixtures of center elements of the color group and, since their number grows exponentially as $3^{volume}$, it is important
to have a way to estimate which critical points are relevant. To this purpose, one may look at the semiclassical approximation

$$Z \approx  \sum_\sigma Z_\sigma^{SC} = (2 \pi)^{\frac{n}{2}} \sum_\sigma n_\sigma \frac{e^{-S(z_\sigma)}}{\sqrt{\prod_i \lambda^{(\sigma)}_i}} \thinspace e^{i\omega_\sigma}
          \equiv (2 \pi)^{\frac{n}{2}} \sum_\sigma n_\sigma e^{-S_{eff}^\sigma} \thinspace e^{i\omega_\sigma}$$

\begin{figure}[tbp]
	\centering
	\includegraphics[scale=0.35]{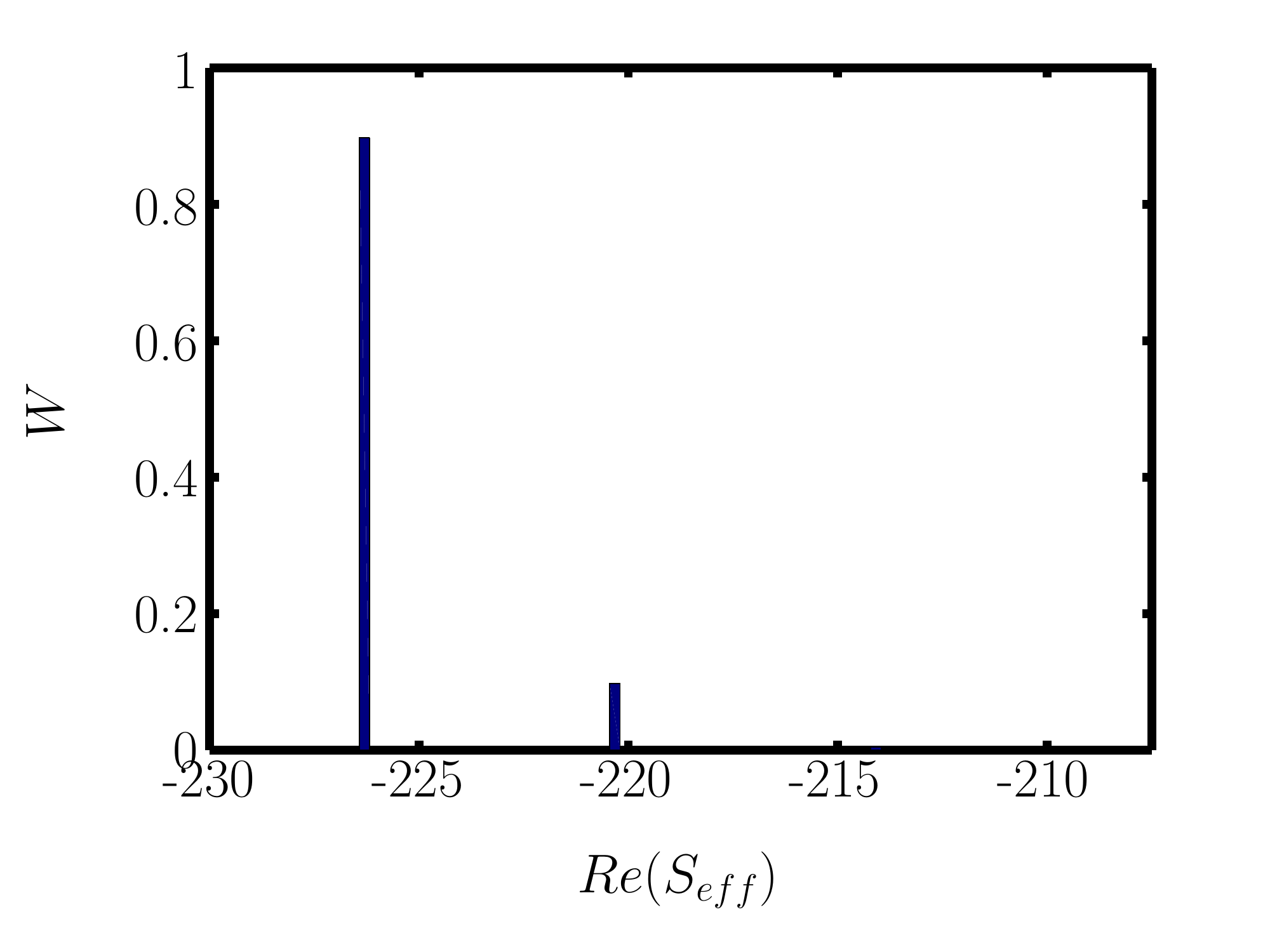}
	\hfill
	\includegraphics[scale=0.35]{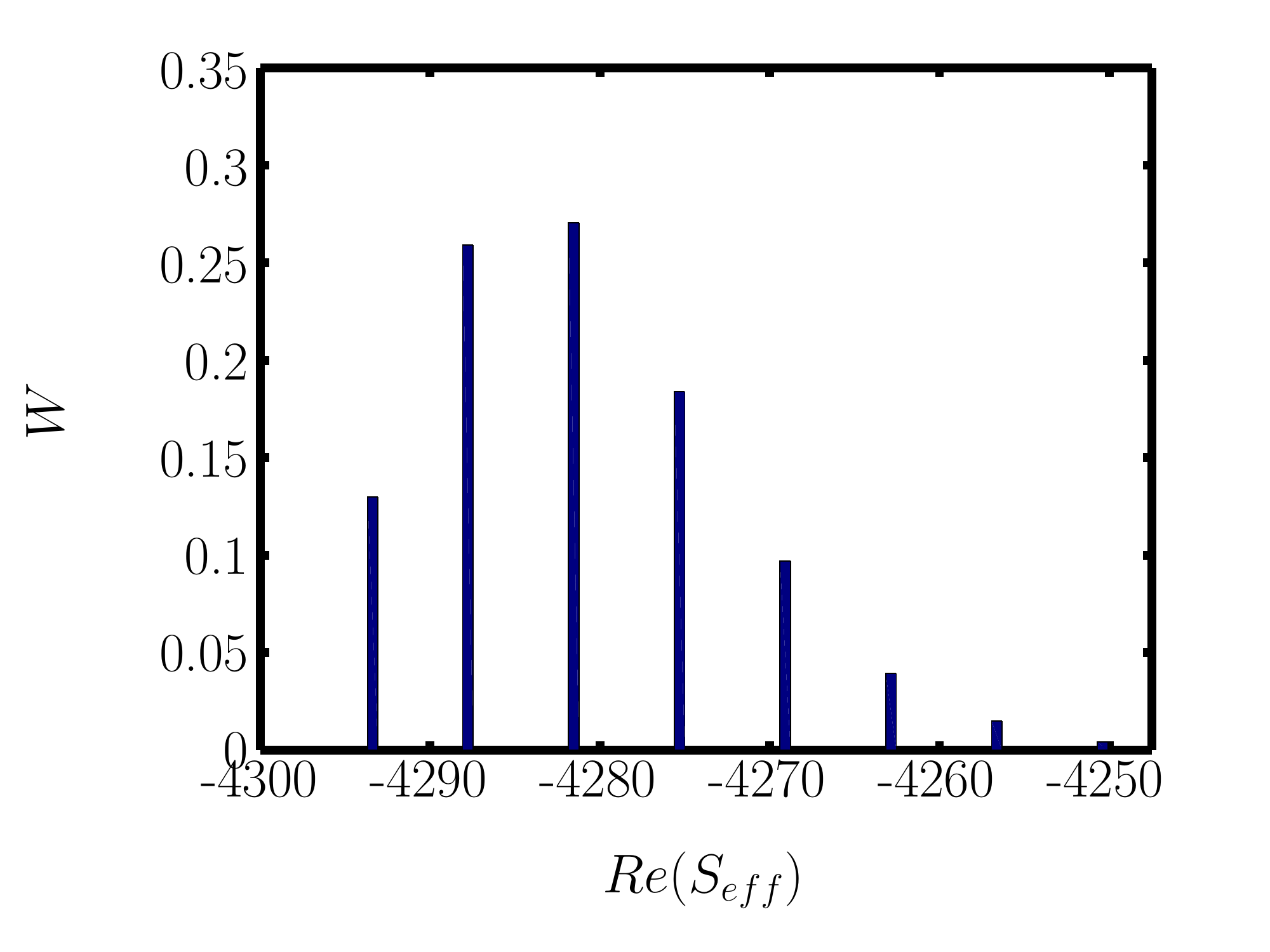}
	\caption{\label{fig:sc1}Semiclassical weights for $V=3^3$ (left) and $V=8^3$ (right).}
\end{figure}

By sampling critical points $\propto e^{-S_{eff}^\sigma}$, one may reconstruct the semiclassical weights of the thimbles attached to such critical points: this amounts to a Monte Carlo
sampling of a spins system. Numerical results are illustrated in fig. \ref{fig:sc1} for two different volumes, with parameters $k = 0.0000887$, $\mu = 0.9990 \thinspace \mu_c$,
$N_t = 116$ and $\beta = 5.7$. The bars located on the $x$ axis correspond to critical points having an increasing number of links set to a center element other than identity and their semiclassical weights (including degeneracy) can be read on the $y$ axis.
For volumes not too large, up to $V \sim 3^3 \div 4 ^3$, only a few critical points are relevant, but as the volume grows this pictures changes significantly and many other critical points become relevant.
In this case, in the thermodynamic limit the fundamental thimble is not the dominant one, which is merely due to combinatorics.


\FloatBarrier


\subsection{Weights determination}
After having identified the relevant critical points, one needs to determine the weights $Z_\sigma$ of the attached thimbles.
In ref. \cite{eruzzi2} it was investigated a region of parameters for one-dimensional QCD in which only three inequivalent thimbles contribute.
Due to a symmetry, two thimbles give contributions that are one the complex conjugate of the other and the thimble decomposition can be put into the form
$$ \langle O \rangle = \frac{n_0 Z_0 e^{-iS_0^I} \langle O e^{i \omega_0}\rangle_0 + n_{12} Z_{12} e^{-iS_{12}^I} \langle O e^{i \omega_{12}}\rangle_{12}}{n_0 Z_0 e^{-iS_0^I} \langle e^{i \omega_0}\rangle_0 + n_{12} Z_{12} e^{-iS_{12}^I} \langle e^{i \omega_{12}}\rangle_{12}} \equiv \frac{\langle O e^{i\omega_0}\rangle_0 + \alpha \langle O e^{i\omega_{12}}\rangle_{12}}{\langle e^{i\omega_0}\rangle_0 + \alpha \langle e^{i\omega_{12}}\rangle_{12}}\mbox{ ,}$$ 

The parameter $\alpha$ was reconstructed by matching the numerical results obtained for one observable to the exact values known from theory
(thus one gives up prediction power for such observable).
While this method worked quite well, it is not entirely satisfactory, as in general many critical points might contribute to the thimble decomposition and one doesn't 
necessarily know the exact values of an equal number of observables beforehand. 

A more general way to determine the weights $Z_\sigma$ by importance sampling can be obtained by observing that in the semiclassical approximation
it is relatively easy to determine both the partial partition function $Z^G_{\hat{n}}$ and the normalized weights $\frac{Z^G_\sigma}{\sum_{\sigma'}Z^G_{\sigma'}}$.
Then, by computing $\frac{Z_\sigma^G}{Z_\sigma} = \int D\hat{n} \frac{Z_{\hat{n}}^G}{Z_{\hat{n}}}\frac{Z_{\hat{n}}}{Z_\sigma} = \langle \frac{Z_{\hat{n}}^G}{Z_{\hat{n}}} \rangle$,
one may recover $\frac{Z_\sigma}{\sum_{\sigma'} Z_{\sigma'}^G}$, which is exactly what is needed up to an irrelevant normalization factor.

\subsection{Numerical results}
We ran one- and two- sites lattice simulations\footnote{The two-sites lattice simulations were finalized in the couple of weeks following the LATTICE2018 event.},
with chemical potential in the range $\mu = 0.9985 \div 1.0015 \thinspace \mu_c$ and parameters $k = 0.0000887$, $N_t = 116$, $\beta = 5.7$
(corresponding to physical parameters $a \approx 0.17$ $fm$, $T \approx 10$ $MeV$, $m_M \approx 20$ $GeV$).
We measured the quark number density and the Polyakov loop, defined as

$$\langle n \rangle = \frac{T}{V} \frac{\partial ln Z}{\partial \mu} = \frac{2}{N_{sites}} \sum_x \frac{h_1 Tr U_x + 2 h_1^2 Tr U_x^{-1} + 3 h_1^3}{1 + h_1 Tr U_x + h_1^2 Tr U_x^{-1} + h_1^3}
\mbox{\hspace{0.5cm}, \hspace{0.5cm}}
\langle L \rangle = \frac{1}{N_{sites}} \sum_x Tr U_x \mbox{ .}$$

\begin{figure}[tbp]
        \centering
        \includegraphics[scale=0.35]{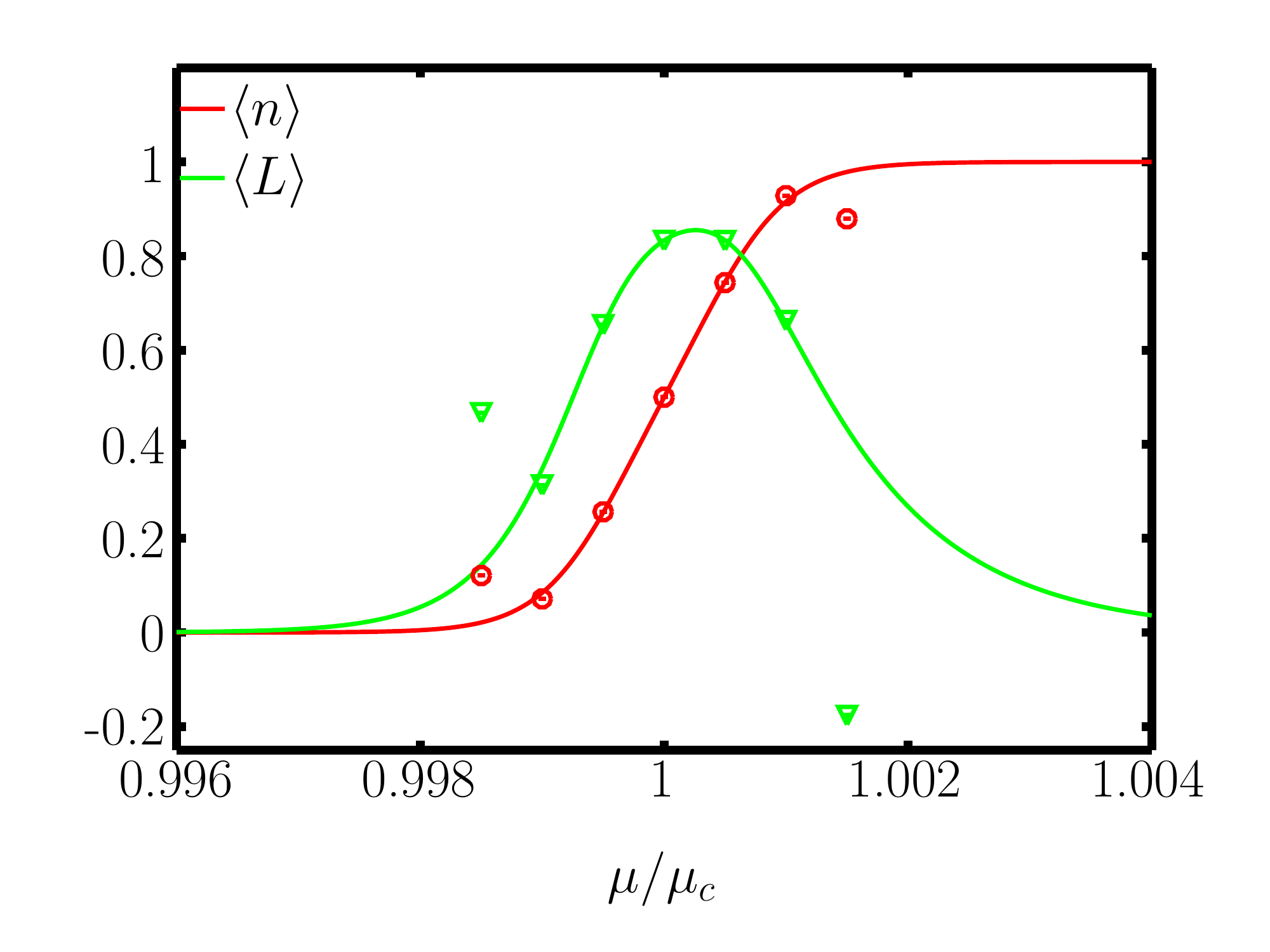}                                                                                                                                           
        \hfill
        \includegraphics[scale=0.35]{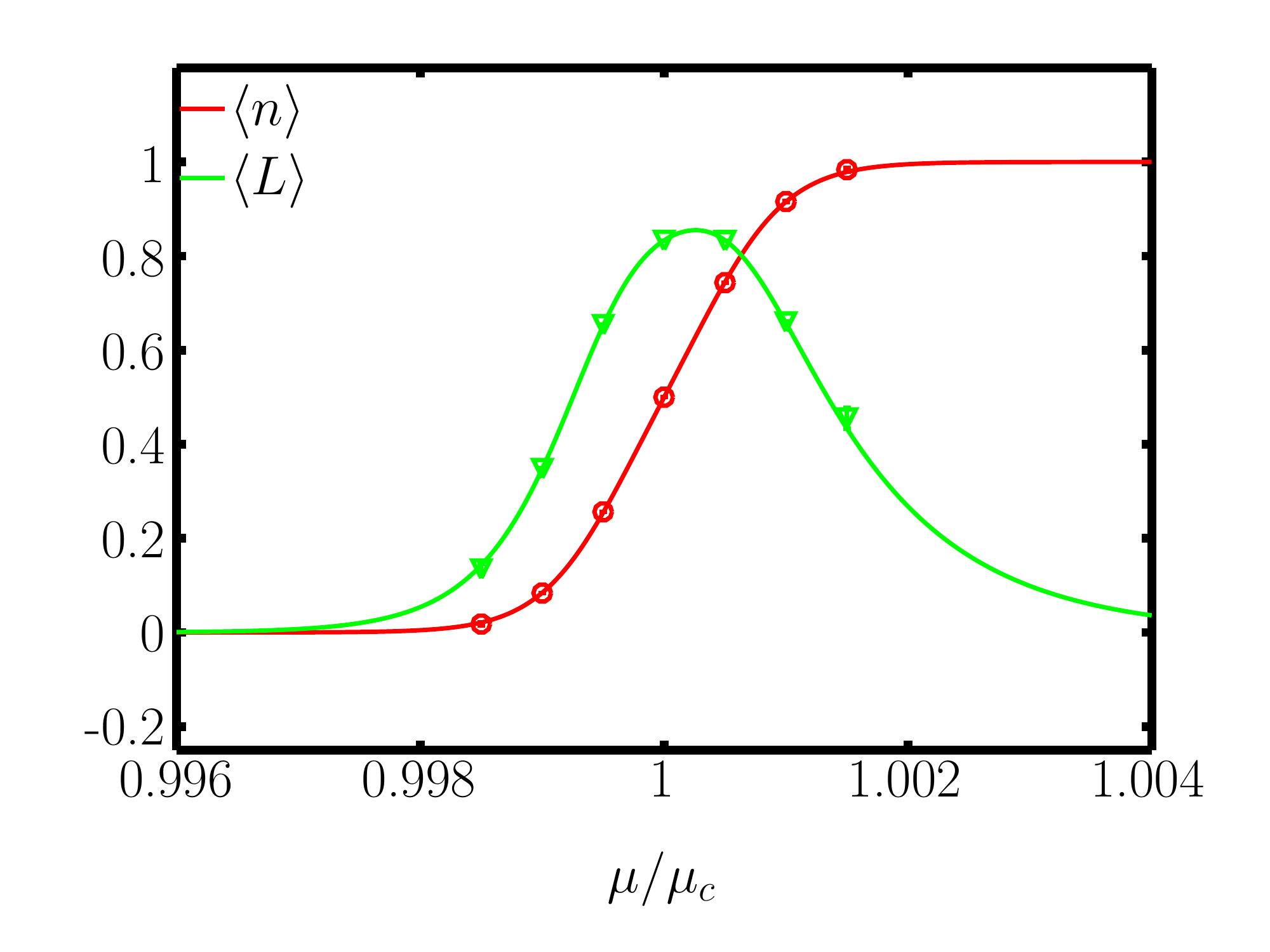}                                                                                                                                           
        \caption{\label{fig:nr1} Results obtained for the quark number density (red) and the Polyakov loop (green) from $1$ thimble (left) and $3$ thimbles (right), $N_{sites}=1$.}
\end{figure}

\begin{figure}[tbp]
        \centering
        \includegraphics[scale=0.35]{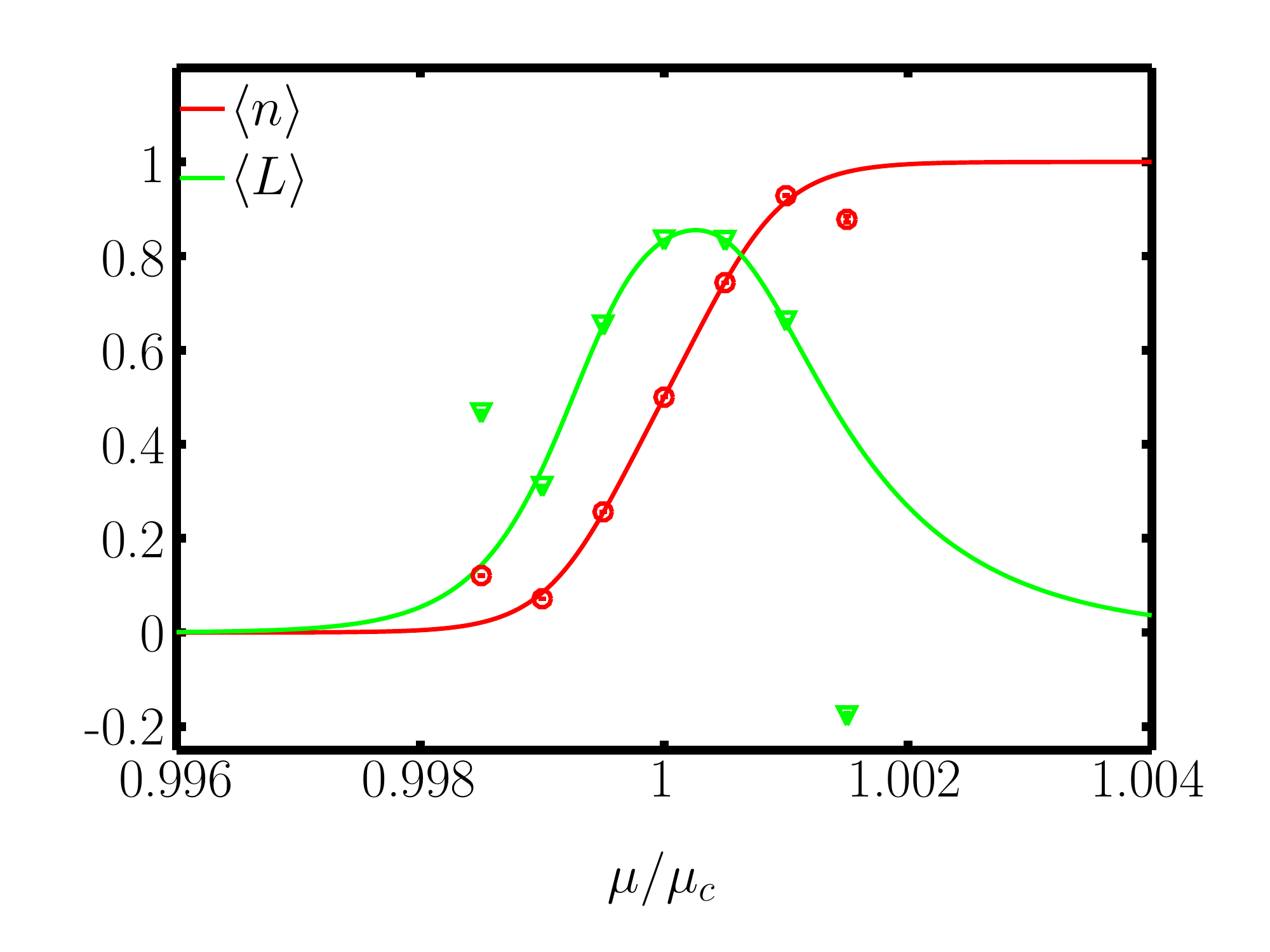}
        \hfill
        \includegraphics[scale=0.35]{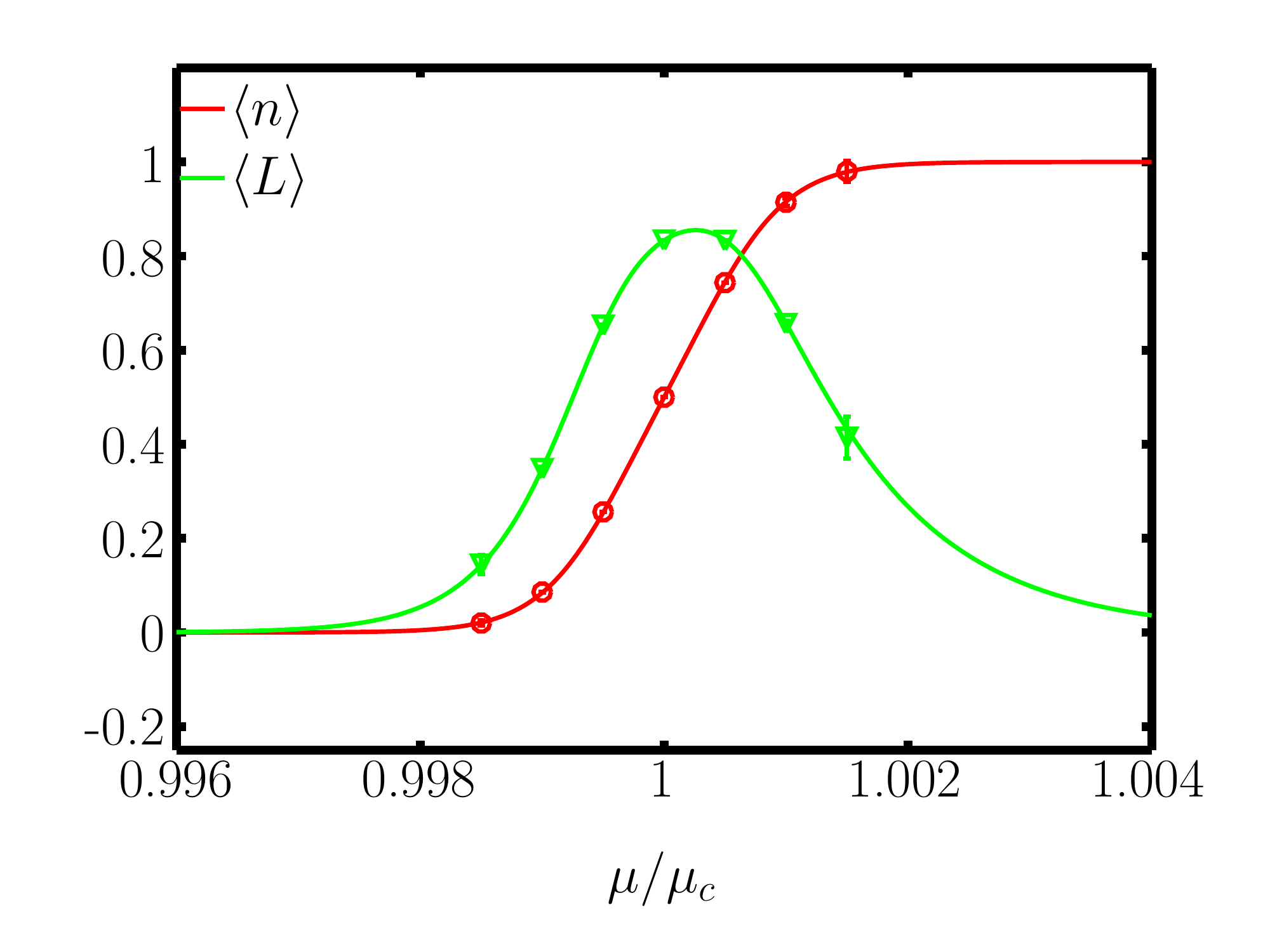}
        \caption{\label{fig:nr2} Results obtained for the quark number density (red) and the Polyakov loop (green) from $1$ thimble (left) and $3$ thimbles (right), $N_{sites}=2$.}
\end{figure}

Numerical results are illustrated in figs. \ref{fig:nr1}-\ref{fig:nr2}: solid lines denote analytic solutions, circles and triangles denote results from simulations.
The results for the quark number density are displayed in red (rescaled by a factor $\frac{1}{6}$), those for the Polyakov loop are displayed in green.
For both lattice volumes, the correct results are only reproduced when taking into account the contributions from three inequivalent thimbles.
The latter are the dominant thimble, associated to the configuration where all links are set to identity, and the next-to-dominant thimbles,
associated to the configurations where all links except one are set to identity.

From the figures it is evident that results are volume independent. Still, they are obtained by summing volume dependent contributions.

\FloatBarrier

\FloatBarrier



\section{Conclusions}
We had a first look at thimble regularization for heavy-dense QCD: by a semiclassical analysis, we have found a region of parameters where a few thimbles contribute for not too
large lattices (up to $\approx 3^3 \div 4^3$ sites). We proposed a first-principle method to determine the weights of the relevant thimbles by importance sampling. We showed that
such method works in one- and two- sites lattice simulations, recovering the correct results from the contributions of three inequivalent thimbles and getting a first glimpse of
the QCD phase diagram using thimbles.

\end{document}